\documentclass[10pt,aps,prb,twocolumn,amsmath,amssymb,superscriptaddress]{revtex4-1}
\usepackage{amsmath}
\usepackage{graphicx}
\usepackage{color}
\usepackage{multirow}
\usepackage[caption = false]{subfig}
\usepackage{array}
\usepackage{mhchem}
\usepackage{mathtools}
\usepackage{siunitx}
\usepackage{tikz}
\tikzset{font={\fontsize{11pt}{12}\selectfont}}
\usetikzlibrary{shapes,arrows}

\makeatletter

\renewcommand*{\p@subsection}{}

\renewcommand*{\p@subsubsection}{}
\makeatother

\usepackage{hyperref}
\hypersetup{
	linkcolor=blue,          % color of internal links (change box color with linkbordercolor)
	citecolor=blue,        % color of links to bibliography
	urlcolor=blue,           % color of external links
}

\def\N{\mathbf{n}}
\def\M{\mathbf{m}}
\def\K{\mathbf{k}}
\def\P{\mathbf{p}}
\def\G{\mathbf{g}}

\newcommand{\ket}[1]{|#1\rangle}

\begin{document}
%\title{Projected Jastrow-Broken-Symmetry-Slater wavefunction using variational Monte Carlo}
%\title{Jastrow correlated and symmetry projected wavefunctions in variational Monte Carlo}
\title{Symmetry Projected Jastrow Mean-field Wavefunction in Variational Monte Carlo}
%\title{Symmetry projected product of Jastrow and broken symmetry mean field wavefunction is variational Monte Carlo}
\author{Ankit Mahajan}
\email{ankit.mahajan@colorado.edu}
\affiliation{Department of Chemistry and Biochemistry, University of Colorado Boulder, Boulder, CO 80302, USA}
\author{Sandeep Sharma}
\email{sanshar@gmail.com}
\affiliation{Department of Chemistry and Biochemistry, University of Colorado Boulder, Boulder, CO 80302, USA}
\begin{abstract}
We extend our low-scaling variational Monte Carlo (VMC) algorithm to optimize the symmetry projected Jastrow mean field (SJMF) wavefunctions. These wavefunctions consist of a symmetry-projected product of a Jastrow and a general broken-symmetry mean field reference. Examples include Jastrow antisymmetrized geminal power (JAGP), Jastrow-Pfaffians, and resonating valence bond (RVB) states among others, all of which can be treated with our algorithm.
%We will demonstrate using benchmark systems including the Nitrogen molecule, a chain of hydrogen atoms, and 2-D Hubbard model that a significant amount of correlation, at a relatively small cost, can be obtained by optimizing the energy of SJMF wavefunction when multiple symmetries including spin, particle number and complex conjugation are simultaneously broken and projected.%
We will demonstrate using benchmark systems including the nitrogen molecule, a chain of hydrogen atoms, and 2-D Hubbard model that a significant amount of correlation can be obtained by optimizing the energy of the SJMF wavefunction. This can be achieved at a relatively small cost when multiple symmetries including spin, particle number, and complex conjugation are simultaneously broken and projected. We also show that reduced density matrices can be calculated using the optimized wavefunctions, which allows us to calculate other observables such as correlation functions and will enable us to embed the VMC algorithm in a complete active space self-consistent field (CASSCF) calculation. %Although symmetry breaking and projection approaches were considered for use in quantum chemistry several years ago, they had not been widely used prior to the work of Scuseria [J. Chem. Theory Comput., 2011, 135, 124108]In the past few years Scuseria \emph{et al}, where they showed that symmetry-projected wavefunctions afford a very accurate mean-field cost representation of strong correlations.
%We will see that projection of certain symmetries is very cheap and natural in VMC in contrast with the deterministic algorithms. We also employ Hilbert space Jastrow correlators to represent long range correlations and suppress spurious fluctuations introduced by symmetry breaking. We will present benchmark results on molecules and the Hubbard model to illustrate the accuracy of this method.
\end{abstract}
\maketitle

\section{Introduction}
Variational Monte Carlo (VMC) is one of the most versatile methods available for obtaining the wavefunction and energy of a system.\cite{RevModPhys.73.33,NigUmr-BOOK-99, TouAssUmr-AQC-15,KolMit-RPP-11,becca2017quantum,Mcmillan,CepCheKal-PRB-77} Compared to deterministic variational methods, VMC allows much greater flexibility in the functional form of the wavefunction. In particular, if one can calculate the overlap of the wavefunction with a Slater determinant at polynomial cost, then it is possible to perform an efficient VMC calculation. Thus sometimes VMC is the only feasible method for calculating wavefunctions of challenging quantum systems.

While the accuracy of VMC is limited by the flexibility of the wavefunction ansatz, projector Monte Carlo (PMC) does not suffer from this shortcoming. However, the cost of performing an unbiased PMC calculation for fermionic systems scales exponentially (with the exception of special cases that display usable symmetries) due to the fermion sign problem. The most common way of overcoming this exponential scaling is by introducing the so-called fixed-node bias. The cost of the fixed-node PMC algorithm is polynomial with the system size, but it no longer delivers unbiased energies. The bias, or the error of PMC, strongly depends on the accuracy of the nodal structure which is often obtained from a VMC calculation. Thus, VMC in addition to being extremely powerful in its own right, also determines the accuracy of various flavors of PMC calculations such as diffusion Monte Carlo (DMC),\cite{GRIMM1971134,Anderson75,CeperleyAlder80} Green's function Monte Carlo (GFMC)\cite{Runge1992,Trivedi1990,Sorella98} and Auxilliary field quantum Monte Carlo (AFQMC).\cite{ZhaKra-PRL-03,Mario-WIREs}

The most commonly used version of VMC is the real-space-VMC, while its orbital-space counterpart is predominately restricted to use with model Hamiltonians such as Hubbard, Heisenberg, etc. The major reason for this is that the cost of performing orbital-space VMC on an \emph{ab-initio} Hamiltonian scales a factor of $O(N)$ worse than both, (1) the cost of performing real-space VMC on an \emph{ab-initio} Hamiltonian and (2) the cost of performing orbital-space VMC on a model Hamiltonian. Recently, we have demonstrated that this cost discrepancy can be removed by introducing three algorithmic improvements\cite{Sabzevari18}. The most significant of these allowed us to efficiently screen the two-electron integrals which are obtained by projecting the \emph{ab-initio} Hamiltonian onto a finite orbital basis. The efficient screening reduced the cost of local energy calculation from $O(N^4)$ to $O(N^2)$ and lowered the overall cost of the algorithm for obtaining a system-size-independent stochastic error from $O(N^5)$ to $O(N^4)$ bringing it on par with other VMC calculations.

In this follow-up work, we will use this newly introduced algorithm to study the performance of the various SJMF states. Projection of symmetry broken mean field (without the Jastrow) wavefunctions is a well-established technique in nuclear physics\cite{ring2004nuclear, blaizot1986quantum} and electronic structure theory.\cite{bach1994generalized,lowdin1955quantum} Recently, these wavefunctions have received renewed attention due to the work of Scuseria \emph{et al}.\cite{scuseria2011projected, jimenez2012projected}, who have shown that several symmetries can simultaneously be broken to recover a significant part of the strong correlation at a mean-field cost. They have also shown  that a greater amount of strong electron correlation can be captured by including a linearized form of the Jastrow factor with these spin-projected reference states.\cite{henderson2013linearized} Attempts to include dynamical correlation in this context have also appeared including perturbation theory,\cite{Takashi} configuration interaction,\cite{Takashi} and coupled cluster.\cite{gus-ccsd}

Unfortunately not all these symmetry projected mean field wavefunctions (e.g., AGP) are size consistent. This shortcoming can be remedied by using local number projectors, which take the form of Hilbert space Jastrow factors as shown by Neuscamman.\cite{neuscamman2012size, neuscamman2013jastrow} In real space VMC, use of Jastrow factors with AGP wavefunctions was first proposed by Sorella \emph{et al}.\cite{casula2003geminal,casula2004correlated,Sorella07} In addition to making the wavefunction size-consistent, the Jastrow correlator also recovers some dynamical correlation missing from the symmetry projected mean field reference.\cite{Neuscamman2016} The drawback is that it is no longer possible to calculate the energy efficiently using a deterministic algorithm and one has to resort to the VMC algorithm. Imada and co-workers\cite{tahara2008variational,tahara2008diff,kurita2015variational,zhao2017variational,darmawan2018stripe} have performed VMC calculations using these SJMF wavefunctions to study model Hamiltonians. Here, we use the full exponential form of the Jastrow on top of a reference that breaks number, spin, and complex conjugation symmetries as the wavefunction ansatz in variational Monte Carlo. We will show that this general wavefunction can be used to study molecular systems across potential energy surfaces and model systems over their parameter space.

The rest of the article is organized as follows. We begin by recapitulating the most important aspects of our VMC algorithm in Section~\ref{sec:algo}. We then discuss the wavefunction ansatz and details of symmetry breaking and projection (\ref{sec:wave}), here we explain the notation used by various researchers and the relations between them. In section \ref{sec:comp}, we present the computational details for evaluating local energy and its gradient efficiently. Finally, we present benchmark results of calculations on the dissociation of the \ce{N2} molecule, hydrogen chain, and the two dimensional Hubbard model (\ref{sec:res}).

\section{Theory and Methods}
\subsection{Algorithm}\label{sec:algo}
In VMC, the energy of a suitably parametrized wavefunction is minimized. The energy of a wavefunction $\Psi(\P)$, where $\P$ is the set of wavefunction parameters, is calculated by performing a Monte Carlo summation.
\begin{align}
\langle H \rangle &= \frac{\sum_{\N} |\langle \Psi(\P) |\N \rangle|^2 \frac{\langle \N|H|\Psi(\P)\rangle}{\langle \N|\Psi(\P)\rangle}}{\sum_{\N} |\langle \Psi(\P)|\N\rangle|^2}\nonumber\\
&= \sum_{\N} \rho_{\N} E_L[\N] = \langle E_L[\N]\rangle_{\rho_{\N}}
\label{eq:vmc}
\end{align}
where, $E_L[\N] = \frac{\langle \N|H|\Psi(\P)\rangle}{\langle \N|\Psi(\P)\rangle}$ is the local energy of a Slater determinant $\N$ and $\rho_{\N} = \frac{ |\langle \Psi(\P) |\N\rangle|^2 }{\sum_{\K} |\langle \Psi(\P)|\K \rangle|^2}$ is the probability distribution of the determinants in the wavefunction. There are three aspects of a VMC algorithm: (1) efficient calculation of the local energy, (2) sample determinants according to the probability distribution $\rho_{\mathbf{n}}$ and (3) optimizing the parameters $\mathbf{p}$ to minimize the energy of the wavefunction. We have recently proposed a set of improvements to all these steps to reduce the cost and lower the scaling of the algorithm. These will be summarized below, but we refer the reader to our original publication Ref.\citenum{Sabzevari18} for more details.

\subsection*{Reduced scaling evaluation of the local energy}
The local energy $E_L[\N]$ of a determinant $|\N\rangle$ is calculated as follows
\begin{align}
 E_L[\N] &=  \frac{\sum_{\M} \langle \N|H|\M\rangle \langle\M|\Psi(\P)\rangle}{\langle \N|\Psi(\P)\rangle} \nonumber\\
 &= \sum_{\M} H_{\N,\M} \frac{ \langle\M|\Psi(\P)\rangle}{\langle \N|\Psi(\P)\rangle},
\end{align}
where the sum is over all determinants $\M$ that have a nonzero transition matrix element ($H_{\N,\M} = \langle \N|H|\M\rangle$) with $\N$. The number of such non-zero matrix elements $H_{\N,\M}$ is on the order of $n_e^2 n_o^2$, where $n_e$ is the number of electrons and $n_o$ is the number of open orbitals. This number increases as a fourth power of the system size and naive use of this formula results in an algorithm that scales poorly with the size of the problem. To reduce the cost of calculating the local energy we truncate the summation over all $\M$ to just a summation over those $\M$, that have a Hamiltonian transition matrix element larger than a user-specified threshold i.e.
\begin{align}
E_L[\N, \epsilon] =&  \sum_{\M}^\epsilon H_{\N,\M} \frac{ \langle\M|\Psi(\P)\rangle}{\langle \N|\Psi(\P)\rangle},\label{eloc:hci}
\end{align}
where $\epsilon$ on the summation denotes that only those terms are included for which $|H_{\N,\M}| > \epsilon$. Note that in the limit that $\epsilon \rightarrow 0$, we recover the exact local energy, $E_L[\N, \epsilon]  \rightarrow E_L[\N]$. It is useful to note that when a local basis set is used the number of elements $H_{\N,\M}$ that have a magnitude larger than a fixed non-zero $\epsilon$ scales quadratically with the size of the system. Thus if we are able to efficiently screen the transition matrix elements for a given $\epsilon \neq 0$, no matter how small the $\epsilon$ is, we are guaranteed to obtain a quadratically scaling evaluation of the local energy. This trick of screening matrix elements is inspired by the heat-bath configuration interaction (HCI) algorithm.\cite{Holmes2016b}

\subsection*{Continuous time Monte Carlo for sampling determinants}
The usual procedure for generating determinants $\N$ according to a probability distribution $\rho_{\N}$ uses the Metropolis-Hastings algorithm in which a random walk is performed to generate a Markov chain.  The efficiency of this algorithm depends on the proposal probability distribution used in simulating the random walks.
A good proposal probability distribution will lead to large steps with very few rejections, but in practice, it is not easy to suggest such a distribution. In this work, we bypass the need for devising complicated proposal probability distributions, by using the continuous time Monte Carlo (CTMC).\cite{BORTZ197510,GILLESPIE1976403} This is an alternative to the Metropolis-Hastings algorithm and has the advantage that every proposed move is accepted. Here, the CTMC algorithm is realized by using the following steps:
\begin{enumerate}
\item Starting from a determinant $\N$ calculate the quantity
\begin{align}
r(\M\leftarrow \N) = \left(\frac{\rho(\M)}{\rho(\N)}\right)^{1/2} = \left|\frac{ \langle\M|\Psi(\P)\rangle}{\langle \N|\Psi(\P)\rangle}\right| \label{eq:rate}
\end{align}
 for all determinants $\M$ that are connected to $\N$ by a single excitation or a double excitation. %(with a Hamiltonian transition matrix element with an absolute value greater than $\epsilon$).
\item Calculate the quantity $t_\N = \frac{1}{\sum_{\M} r(\M\leftarrow \N) }$ and assign that weight to the walker $\mathbf{n}$.
\item Next, a new determinant is selected, without rejection, out of all the determinants $\M$ with a probability proportional to $r(\M\leftarrow \N)$.
\end{enumerate}
We note that in the VMC algorithm, the quantities $\left|\frac{ \langle\M|\Psi(\P)\rangle}{\langle \N|\Psi(\P)\rangle}\right|$ are already used in the calculation of the local energy (see Equation~\ref{eloc:hci}) and just by storing those quantities the CTMC algorithm can be used with almost no overhead once the local energy has been calculated.

\subsection*{AMSGrad algorithm for optimizing the energy}
The optimized wavefunction ($\Psi(\P)$) is obtained by minimizing its energy with respect to its parameters $\P$. This is a challenging optimization problem because the energy is a non-linear function of the wavefunction parameters. Further, the gradient of the energy with respect to the parameters is noisy because a stochastic method is used. Several first-order optimization algorithms (that only require gradients) such as the conjugate gradient method become unstable when the gradient is noisy. Booth and coworkers\cite{PhysRevLett.118.176403} first proposed the use of adaptive stochastic gradient descent (SGD) optimization in VMC. In our previous work, we have demonstrated that the SGD method called AMSGrad\cite{Reddi2017} can be used to effectively optimize the VMC wavefunctions. In AMSGrad an exponentially decaying moving average of the first and second moments $\mathbf{m}$ and $\mathbf{n}$ are respectively calculated
\begin{align}
\mathbf{m}^{(i)} =& (1-\beta_1)  \mathbf{m}^{(i-1)} + \beta_1\G^{(i)}\label{eq:m1}\\
\mathbf{n}^{(i)} =& \max(\mathbf{n}^{(i-1)}, (1-\beta_2) \mathbf{n}^{(i-1)} +  \beta_2(\G^{(i)}\cdot\G^{(i)})\label{eq:m2}
\end{align}
where, $\beta_1$ and $\beta_2$ are used to determine the rate of decay. These first and second moments are then used to update the parameters ($\mathbf{p}$)
\begin{align}
\Delta\P_j =& -\alpha \mathbf{m}_j^{(i)}/\sqrt{\mathbf{n}^{(i)}_j}\label{eq:m3}
\end{align}
where, $\alpha$ determines the step size. In equations~\ref{eq:m2} and ~\ref{eq:m3}, the product and division are element-wise operations.
AMSgrad has the advantage that the CPU and memory cost scales linearly with the number of wavefunction parameters and in our tests, it outperformed simple gradient descent. In the calculations performed in this paper, we have used the parameters $\alpha = 0.01, \beta_1 = 0.1, \beta_2=0.01$ which give satisfactory convergence rates (in some cases we have to run a few iterations with less aggressive parameters until reasonable estimates of the first and second moments $\mathbf{m}$ and $\mathbf{n}$ are built up).

\subsection{Wavefunctions}\label{sec:wave}
The accuracy of the VMC results depends critically on the wavefunction ansatz employed. The ansatz needs to be general enough to capture the relevant physics of the system, however, to be computationally tractable with the VMC algorithm, the wavefunction must allow efficient computation (at polynomial cost) of the overlap with a Slater determinant. A wavefunction that satisfies both these requirements is used in this work and has the form
\begin{equation}
   \ket{\psi} = \hat{C}\hat{P}\ket{\phi_0},
\end{equation}
where $\hat{C}$ is a correlator and $\hat{P}$ is the projector that restores symmetries of the symmetry-broken mean-field reference $\ket{\phi_0}$. A combination of different correlators and references can be used to represent the ground state of the system under study. In this section, we study each of these terms in detail.

\subsection*{Mean field Reference}
The reference describes uncorrelated electrons, in other words, it is the ground state of a general quadratic Hamiltonian. Here we will allow the mean field wavefunction ($\phi_0$) of the system to break the symmetries of the Hamiltonian, which gives the wavefunction additional variational flexibility resulting in lower energies. However, the wavefunction also has a projector ($\hat{P}$), that restores these symmetries. The functional form of the resulting wavefunction $\hat{P}|\phi_0\rangle$ depends on the symmetries that are broken and restored, which we will describe in this section.

%\subsubsection*{Symmetry breaking and projection}
In a finite system, the true ground state obeys all the symmetries of the Hamiltonian. On the other hand, the VMC wavefunction is an approximate ansatz and forcing it to obey these symmetries can only restrict its variational freedom thereby raising its energy.\cite{lykos1963discussion} One can get around this by allowing the reference to break the symmetries and then projecting it onto the desired symmetry sector. This has the advantage of affording the wavefunction more variational freedom as well as correct symmetries. More physically, breaking symmetries introduces quantum fluctuations in the reference necessary for representing multideterminant states while projection serves to filter out unwanted fluctuations.

We note an important point the regarding optimization of such wavefunctions. One could either optimize the symmetry-broken reference without projectors and apply the projectors after optimization. Alternatively, one could optimize the symmetry-broken reference in the presence of projectors. The wavefunction produced by the former procedure is in the variational space explored by the latter. Thus the variation after projection approach is more general and always gives lower energies. In this work, we will use the wavefunction obtained by performing the variation after projection approach.

Here we will break and restore the particle number, spin, and complex conjugation symmetries. Molecular electronic systems in the absence of spin-orbit coupling and Hubbard model with real hopping parameters obey all these symmetries, i.e.
 \begin{equation*}
   \left[\hat{H}, \hat{N}\right] = \left[\hat{H}, \mathbf{\hat{S}}\right] = \left[\hat{H}, \hat{K}\right] = 0,
\end{equation*}
where $\hat{N}$ is the number operator, $\mathbf{\hat{S}}$ is the vector spin operator, and $\hat{K}$ is the complex conjugation operator. The number ($U(1)$) and spin ($SU(2)$) symmetries are continuous, while complex conjugation is discrete. The projection after variation approach has previously been used in deterministic algorithms,\cite{scuseria2011projected, jimenez2012projected} where continuous symmetry projectors were written by discretizing the integrals obtained by using the generator coordinate method, while the discrete projector $\hat{P}_K$ was implemented by diagonalizing a $2\times 2$ matrix. In VMC, these expensive integrals can be avoided for certain symmetries. To see this, recall that the central quantity of interest is the overlap ($\langle \mathbf{n}|\hat{P}|\phi_0\rangle$) of the wavefunction ($\hat{P}|\phi_0\rangle$) with a walker which is simply a Slater determinant ($|\mathbf{n}\rangle$). In VMC, instead of applying the projector on to the symmetry broken mean-field wavefunction, we apply it to the walker $|\mathbf{n}\rangle$. Thus, the $\hat{N}$ and $\hat{S}_z$ projections can be done by using walkers with the desired number of electrons and spin component. The $\hat{S}^2$ projection still needs to be done using an integral projector and we will not be using it here. Complex conjugation symmetry can be restored by simply taking the real part of the overlap ($\langle \mathbf{n}|\hat{P}|\phi_0\rangle$).

%\subsubsection*{Slater determinants and pairing wavefunctions}

First, let's consider reference states that have a fixed particle number, i.e. those obeying the particle number symmetry. These are the Slater determinants widely used in Hartree-Fock (HF) methods. The most general form of a Slater determinant used is the generalized HF (GHF) which is given by
\begin{equation}
   \ket{\text{GHF}} = \prod_{k=1}^{N_e}\hat{a}_k^{\dagger}\ket{0},
\end{equation}
where $N_e$ is the number of electrons and $\hat{a}_k^{\dagger}$ creates an electron in the molecular orbital $k$ given by
\begin{equation}
   \hat{a}_k^{\dagger} = \sum_{p=1}^{M}\sum_{\sigma=\uparrow,\downarrow}\theta^{p\sigma}_{k}\hat{a}_{p\sigma}^{\dagger}\label{eq:ref},
\end{equation}
where $\hat{a}_{p\sigma}^{\dagger}$ creates an electron in the spatially local atomic orbital $p$ with spin $\sigma$, $M$ is the number of atomic orbitals, and $\theta^{p\sigma}_{k}$ are complex numbers. In this paper, we order the spin orbital indices such that all spin up orbitals come before all the spin down ones, i.e. $i\uparrow < j\downarrow$ for all $i$ and $j$. Note that the GHF molecular orbitals are not separable into spatial and spin parts in general, i.e. their spatial and spin degrees of freedom can be entangled. By putting constraints on the $\theta$ matrix we can obtain specialized forms of Slater determinants. The unrestricted Hartree Fock (UHF) wavefunction is given by
\begin{equation}
   \ket{\text{UHF}} = \prod_{k=1}^{N_{\uparrow}}\hat{a}_{k\uparrow}^{\dagger}\prod_{l=1}^{N_{\downarrow}}\hat{a}_{l\downarrow}^{\dagger}\ket{0},
\end{equation}
where
\begin{equation*}
   \hat{a}_{k\uparrow}^{\dagger} = \sum_{p=1}^{M}\theta^{p}_{k\uparrow}\hat{a}_{p\uparrow}^{\dagger},\qquad  \hat{a}_{l\downarrow}^{\dagger} = \sum_{p=1}^{M}\theta^{p}_{l\downarrow}\hat{a}_{p\downarrow}^{\dagger}.
\end{equation*}
In restricted Hartree Fock (RHF), the state is further restricted by the requirement $\theta^{p}_{k\uparrow}=\theta^{p}_{k\downarrow}$. %Note that RHF can only be used for unpolarized systems ($N_{\downarrow}=N_{\uparrow}$).

Now, let's look at reference states that break the particle number symmetry. Here we will only consider systems with an even number of electrons, although extension to the odd case is possible. The most general such state for a system with even number of electrons is given by
\begin{equation}
   \ket{\text{GBCS}} = \exp \left(\sum_{p\sigma,q\gamma} F_{p\sigma,q\gamma}\hat{a}_{p\sigma}^{\dagger}\hat{a}_{q\gamma}^{\dagger}\right)\ket{0},
\end{equation}
where $p$ and $q$ are the spatial orbital indices, while $\sigma$ and $\gamma$ are spin indices. $F_{p\sigma,q\gamma}$ are complex numbers and $F_{p\sigma,q\gamma}=-F_{q\gamma,p\sigma}$ due to fermionic anticommutation. This is a generalized Bardeen-Cooper-Schrieffer (GBCS) wavefunction. Its number projected form ($\hat{P}_N\ket{\text{GBCS}}$) is known as Pfaffian\cite{becca2017quantum}
 \begin{equation}
   \ket{\text{Pf}} = \hat{P}_N\ket{\text{GBCS}} =\dfrac{1}{p!}\left[\sum_{m\sigma,n\gamma}F_{m\sigma,n\gamma}\hat{a}_{m\sigma}^{\dagger}\hat{a}_{n\gamma}^{\dagger}\right]^{N_e/2}\ket{0}.\label{eq:pf}
\end{equation}
By allowing only opposite spin triplet pairings, i.e.
\begin{equation}
   \ket{\text{UBCS}} = \exp \left(\sum_{p,q} F_{p\uparrow,q\downarrow}\hat{a}_{p\uparrow}^{\dagger}\hat{a}_{q\downarrow}^{\dagger}\right)\ket{0},
\end{equation}
we get the unrestricted BCS (UBCS) wavefunction. Further restricting the pairing matrix to be symmetric ($F_{p\uparrow,q\downarrow} = F_{q\uparrow,p\downarrow}$) ensuring that each pairing is a singlet, we get the conventional restricted BCS (RBCS) wavefunction. Its number projected form ($\hat{P}_N\ket{\text{RBCS}}$) is known as antisymmetrized geminal product (AGP).\cite{casula2003geminal,casula2004correlated}

Despite their distinct appearance, Slater determinants and pairing wavefunctions are closely related\cite{misawa2019mvmc}. We can express the GHF wavefunction in a pairing form as follows:
 \begin{equation*}
   \begin{split}
       \ket{\text{GHF}} &= \prod_{k=1}^{N_e}\hat{a}_k^{\dagger}\ket{0}\\
			 &= \left[\prod_{p=1}^{N_e/2}\left(\hat{a}_{2p-1}^{\dagger}\hat{a}_{2p}^{\dagger}\right)\right]\ket{0}\\
			 &\propto \left(\sum_{p=1}^{N_e/2}\hat{a}_{2p-1}^{\dagger}\hat{a}_{2p}^{\dagger}\right)^{N_e/2}\ket{0}\\
			 &= \left[\sum_{m\sigma,n\gamma}\left(\sum_{p=1}^{N_e/2}\theta^{m\sigma}_{2p-1}\theta^{n\gamma}_{2p}\right)\hat{a}_{m\sigma}^{\dagger}\hat{a}_{n\gamma}^{\dagger}\right]^{N_e/2}\ket{0},\\
   \end{split}
\end{equation*}
where in the third line we have ignored an unimportant normalization factor and used the fact that products of pairs of creation operators commute with each other. This shows that GHF is a special form of Pfaffians given in Equation~\ref{eq:pf}. The explicit relation between the GHF coefficient matrix and the corresponding Pfaffian pairing matrix is thus given (to within an unimportant overall normalization factor) by
\begin{equation}
	F = \theta A \theta^T,
\end{equation}
where $A$ is a $N_e \times N_e$ block diagonal matrix with $\begin{bmatrix} 0 & 1\\ -1 & 0 \end{bmatrix}$ as blocks. We can similarly prove that RHF is a special case of AGP.
\begin{table}
\caption{The table lists the various mean-field wavefunctions and the symmetries that are broken in them.}
\centering
\begin{tabular}{|c|c|}
\hline
Ansatz & Broken symmetries\\
\hline
GHF & $S_z$, $S^2$, $K$\\
UHF & $S^2$, $K$\\
RHF & $K$\\
GBCS &  $N$, $S_z$, $S^2$, $K$\\
UBCS & $N$, $S^2$, $K$ \\
RBCS & $N$, $K$ \\
\hline
\end{tabular}
\end{table}

%{\color{red}\subsection*{Properties}
%\begin{itemize}
%	\item Size consistency
%	\item Size extensivity
%	\item Correlation and pair functions
%	\item Seniority number: Multideterminant expansions
%\end{itemize}}

\subsection*{Correlators}
%Electrons interact with each other via the long ranged Coulomb potential. This leads to
Correlation between the motion of different electrons is not completely captured by the reference outlined above. The correlators ($\hat{C}$) acting upon a reference encode these correlations explicitly and can in principle, with sufficiently large correlators, account for 100\% of the electron correlation. We have previously used correlated product states (CPS)\cite{Neuscamman2011,Neuscamman2012} as correlators, however, here we use Hilbert space Jastrows because of their more compact representation. They are completely equivalent to two-electron CPS. We use the following form of the Jastrow:
\begin{align}
   \hat{\mathcal{J}} &= \exp\left(\sum_{p\sigma\geq q\gamma} v_{p\sigma,q\gamma}\hat{n}_{p\sigma}\hat{n}_{q\gamma}\right) \nonumber\\
   &= \prod_{p\sigma\geq q\gamma} \left[1 - \hat{n}_{p\sigma}\hat{n}_{q\gamma} \left(1-J_{p\sigma,q\gamma}\right)\right]\label{eq:jastrow},
\end{align}
where $n_{p\sigma}$ and $n_{q\gamma}$ are number operators for the spin orbitals $p\sigma$ and $q\gamma$, respectively and $J_{p\sigma,q\gamma}$ are the variational parameters related to $v_{p\sigma,q\gamma}$ in the exponential form by
\begin{equation*}
    J_{p\sigma,q\gamma} = \exp(v_{p\sigma,q\gamma}).
\end{equation*}
The Jastrow is not invariant to unitary rotations of these spin orbitals and thus a judicious choice is necessary to ensure good quality. Although the optimal choice of the spin orbital basis is far from obvious, the use of local basis ensures that the wavefunction is size consistent due to its ability to perform local particle number projections.\cite{neuscamman2012size, neuscamman2013jastrow, neuscamman2016subtractive} Jastrows in local basis includes the onsite Gutzwiller factors\cite{gutzwiller} as well as long-range density correlations.\cite{capello2005luttinger,capello2005variational} Thus, in this work, we use local bases to represent the Jastrows. It has also been shown that Jastrow is a limited form of coupled cluster doubles operator,\cite{Neuscamman2016} that impart some dynamical correlation to the wavefunction. %Although they are cheaper than coupled cluster, they are also not as effective at adding correlation.

\subsection{Computational details}\label{sec:comp}
At each iteration of the VMC algorithm, the overlap between a walker and the wavefunction is needed. This overlap can be calculated by using the expression
 \begin{equation}
       \langle\M|\psi\rangle = \langle\M|\hat{C}\hat{P}|\phi_0\rangle =C\left[\M\right]\langle\M|\hat{P}|\phi_0\rangle,
\end{equation}
where we have used the fact that the Jastrow is diagonal in the configuration space of the local orbitals.
$\ket{\M}$ is a determinant in the local orbital Hilbert space used in the definition of references orbitals (Eq~\ref{eq:ref}) and Jastrow factors (Eq.~\ref{eq:jastrow}). Let us examine each of these terms in more detail.

\begin{itemize}
\item  The overlap with the Jastrow is simply given by
\begin{equation}
   C[\M] = \prod_{p\sigma,q\gamma}^{\text{occ.}}J_{p\sigma,q\gamma},
\end{equation}
where the product is over all pairs of occupied spin orbitals. This computation has $O(N^2)$ cost.

\item For the projectors considered here, we have
 \begin{equation}
		 \langle\M|\hat{P}_K\hat{P}_{S_z}\hat{P}_N|\phi_0\rangle = \text{Re}\left[\langle\M|\phi_0\rangle\right],
\end{equation}
where we have used the fact that the walker $\ket{\M}$ is an $\hat{S_z}$ and $\hat{N}$ eigenstate with desired eigenvalues.

\item When $|\phi_0\rangle$ is the GHF state, we get
\begin{equation*}
   \langle\M|\text{GHF}\rangle = \text{det}(\theta[\M]),
\end{equation*}
where $\text{det}(\theta[\M])$ is the determinant of the $N\times N$ matrix $\theta\left[\M\right]$ which itself is the slice of the coefficient matrix obtained by using the rows corresponding to spin orbitals occupied in $\ket{\M}$. Overlaps can be calculated similarly for UHF and RHF. For the GBCS wavefunction, we have
\begin{equation}
   \langle\M|\text{GBCS}\rangle = \langle\M|\text{PF}\rangle = \text{pf}(F[\M]),
\end{equation}
where $F \left[\M\right]$ is the slice of the pairing matrix obtained by using rows and columns corresponding to spin orbitals occupied in  $\ket{\M}$. Pfaffian of a skew-symmetric matrix $[A]_{2M\times 2M}$ is defined as
\begin{equation}
   \text{pf}(A) = \sum_{\mathcal{P}}(-1)^{\mathcal{P}}A_{i_1j_1}A_{i_2j_2}\dots A_{i_Mj_M},
\end{equation}
where the sum is over all partitions $\{(i_k, j_k)\}$ of the $2M$ indices with $i_k<j_k$ and $(-1)^{\mathcal{P}}$ is the parity of the partition $\mathcal{P}$. It is possible to calculate the Pfaffian of a skew-symmetric matrix in $O(N^3)$ (same as the determinant calculation) steps using the Parlett-Reid algorithm which is based on Gaussian transformations.\cite{wimmer2012algorithm} Pfaffian has the property
\begin{equation*}
   \text{pf}\begin{bmatrix} 0 & C \\ -C^T & 0 \end{bmatrix} \propto \text{det}(C).
\end{equation*}
The pairing matrix for BCS has the form on the left side of the above equation. Thus
\begin{equation}
   \langle\M|\text{BCS}\rangle = \langle\M|\text{AGP}\rangle = \text{det}(F[\M]),
\end{equation}
where $F \left[\M\right]$ is the slice of the AGP pairing matrix with rows and columns corresponding to up and down spin orbitals occupied in $\ket{\M}$, respectively.
\end{itemize}

\subsection*{Local energy calculation}
The local energy $E_L[\N]$ of a determinant $|\N\rangle$ is calculated as follows
\begin{equation}
 E_L[\N] =  \frac{\sum_{\M} \langle \N|H|\M\rangle \langle\M|\psi\rangle}{\langle \N|\psi\rangle} = \sum_{\M} H_{\N,\M} \frac{ \langle\M|\psi\rangle}{\langle \N|\psi\rangle},
\end{equation}
where the sum is over all determinants $\M$ that have a nonzero transition matrix element ($H_{\N,\M} = \langle \N|H|\M\rangle$) with $\N$. Note that for the molecular Hamiltonian only determinants connected by at most two electron excitations have a nonzero transition matrix element. For performing fast VMC calculations it is essential to be able to calculate the overlap ratios
\begin{equation}
   \frac{\langle\M|\psi\rangle}{\langle \N|\psi\rangle} = \dfrac{C[\M]}{C[\N]}\dfrac{\langle\M|\hat{P}|\phi_0\rangle}{\langle\N|\hat{P}|\phi_0\rangle}.
\end{equation}
efficiently.
A naive calculation of the correlator overlap ratio by individually calculating both numerator and denominator has cost $O(N^2)$. We can reduce this cost by calculating and storing the following vector at the start of the calculation:
\begin{equation*}
   v_{\N}[i\sigma] = J_{i\sigma,i\sigma}\prod_{p\gamma \in \text{ occ.}}^{p\gamma \neq i\sigma}J_{i\sigma,p\gamma},
\end{equation*}
which has a length equal to the number of local spin orbitals. The overlap ratio with a determinant $\ket{\M}$ obtained from $\ket{\N}$ by a single excitation $i\sigma \rightarrow a\sigma$ is given in terms of $v_{\N}$ as
\begin{equation*}
   \dfrac{C[\M]}{C[\N]} = J_{i\sigma,a\sigma}\dfrac{v_{\N}[a]}{v_{\N}[i]}.
\end{equation*}
A similar equation, but still with $O(1)$ cost, can be obtained for double excitations. As the walker moves during a simulation, changing by at most two excitations, the $v_{\N}$ vector is updated in $O(N)$ steps.

A naive implementation of the overlap ratios for the reference would have an $O(N^3)$ cost. Here also, we can store intermediate quantities and reduce this cost by virtue of the fact that the determinants in the ratio differ by at most a double excitation.
\begin{align}
   \dfrac{\langle\M|\hat{P}|\phi_0\rangle}{\langle\N|\hat{P}|\phi_0\rangle} &= \dfrac{\text{Re}\left[\langle\M|\phi_0\rangle\right]}{\text{Re}\left[\langle\N|\phi_0\rangle\right]}\\
   &= \text{Re}\left[\dfrac{\langle\M|\phi_0\rangle}{\langle\N|\phi_0\rangle}\langle\N|\phi_0\rangle\right]\bigg/\text{Re} \left[\langle\N|\phi_0\rangle\right]
\end{align}
For GHF Slater determinants, we have
\begin{equation}
   \dfrac{\langle\M|\text{GHF}\rangle}{\langle\N|\text{GHF}\rangle} = \dfrac{\text{det}(\theta[\M])}{\text{det}(\theta[\N])}.
\end{equation}
An excitation of one electron amounts to a change of a row in the determinant. For a determinant $\ket{\M}$ obtained from the excitation $i\sigma \rightarrow a\sigma$, the ratio can be calculated using the Woodbury lemma\footnote{M. Brookes, The Matrix Reference Manual, (2011); see online at http://www.ee.imperial.ac.uk/hp/staff/dmb/matrix/intro.html.} :
\begin{equation}
   \dfrac{\text{det}(\theta[\M])}{\text{det}(\theta[\N])} = \sum_{p\gamma} \theta_{a\sigma,p\gamma}(\theta[\N]^{-1})_{p\gamma,i\sigma} = R[\N]_{a\sigma,i\sigma}.
\end{equation}
% Note that in this and subsequent such expressions (containing matrix slices dependent on the walker) the row and column indices refer to the full matrix and need to be converted into the relative indices of the matrix slice.
Thus, by precalculating and storing the matrix $R[\N]_{a\sigma,i\sigma}$, the ratios of determinants can be calculated with an $O(1)$ cost. Similar expressions can be derived for double excitations that allow calculations of overlap ratios to be evaluated at $O(1)$ cost. Calculation and update of the $R[\N]_{a\sigma,i\sigma}$ matrix require the inverse of $\theta[\N]$. We can avoid calculating the inverse at every step, an $O(N^3)$ scaling operation, by updating it using the Sherman-Morrison formula in $O(N^2)$ steps. Similar relations can also be derived for RHF and UHF references. As outlined earlier, the AGP overlaps are given by determinants as well which implies that their overlap ratios can be calculated using similar manipulations with an $O(N^2)$ cost.

Overlap ratio for the GBCS wavefunction is given by\cite{bajdich2006pfaffian,bajdich2008pfaffian}
\begin{equation}
   \dfrac{\langle\M|\text{GBCS}\rangle}{\langle\N|\text{GBCS}\rangle} = \dfrac{\text{pf}(F[\M])}{\text{pf}(F[\N])}.
\end{equation}
Again instead of calculating both overlaps separately, the ratio can be calculated at a reduced cost by using an identity analogous to the determinant lemma, given by\cite{becca2017quantum, morita2015quantum, misawa2019mvmc}
\begin{equation*}
	\begin{split}
		F[\M] &= F[\N] + BCB^T\\
   \dfrac{\text{pf}(F[\M])}{\text{pf}(F[\N])} &= \dfrac{\text{pf}(C^{-1}+B^TF[\N]^{-1}B)}{\text{pf}(C^{-1})},\\
 \end{split}
\end{equation*}
where $C$ is a $2m\times 2m$ skew-symmetric matrix and $B$ is $N\times2m$ matrix, chosen to affect the update in the top equation for an $m$ electron excitation. Since $m$ is at most 2, the Pfaffians on the right hand side can be calculated explicitly. For a single excitation $i\sigma \rightarrow a\sigma$ by choosing the $B$ and $C$ matrices appropriately, we get
\begin{equation}
   \dfrac{\text{pf}(F[\M])}{\text{pf}(F[\N])} = (F^c[\N]F[\N]^{-1})_{a\sigma,i\sigma} = R[\N]_{a\sigma,i\sigma},
\end{equation}
where $F^c[\N]$ is the slice of the pairing matrix obtained by using rows and columns corresponding to unoccupied and occupied spin orbitals in $\ket{\N}$, respectively. For a double excitation $i\sigma \rightarrow a\sigma$ and $j\gamma \rightarrow b\gamma$ (assuming $i\sigma < j\gamma$), we get
\begin{align}
   \dfrac{\text{pf}(F[\M])}{\text{pf}(F[\N])}  =& F[\N]^{-1}_{i\sigma,j\gamma}\left[(F^c[\N]F[\N]^{-1}F^r[\N])_{b\gamma,a\sigma} + F_{b\gamma,a\sigma}\right] \nonumber \\
   &+ R[\N]_{a\sigma,i\sigma}R[\N]_{b\gamma,j\gamma}-R[\N]_{b\gamma,i\gamma}R[\N]_{a\sigma,j\gamma},
\end{align}
where $F^r[\N]$ is the slice of the pairing matrix obtained by using rows and columns corresponding to occupied and unoccupied spin orbitals in $\ket{\N}$, respectively. We precalculate the $R[\N]$ and $F^c[\N]F[\N]^{-1}F^r[\N]$ matrices, and update them before each Monte Carlo iteration in $O(N^2)$ cost. To avoid the expensive direct calculation of the inverse in this expression, we instead use the inverse update identity
\begin{align*}
   F[\M]^{-1} = &F[\N]^{-1}\\
   &-  F[\N]^{-1}B(C^{-1}+B^{T}F[\N]^{-1}B)^{-1}B^{T}F[\N]^{-1}.
\end{align*}

\subsection*{Gradient overlap ratios}
Gradient overlap ratios are given by
\begin{equation*}
   \dfrac{\langle\N|\psi_i(\P)\rangle}{\langle\N|\psi(\P)\rangle}=\dfrac{\partial_i\langle\N|\psi(\P)\rangle}{\langle\N|\psi(\P)\rangle},
\end{equation*}
where $\P$ denotes the vector of all wavefunction parameters and
\begin{equation*}
   \ket{\psi_i(\P)} = \bigg\vert\dfrac{\partial\psi(\P)}{\partial\P_i}\bigg\rangle
\end{equation*}
is the wavefunction derivative with respect to the $i$th parameter.
For Jastrow parameters, we have
 \begin{align}
    \partial_{J_{p,q}}\langle\N|\psi(\P)\rangle &= \partial_{J_{p,q}}\left(C[\N]\langle\N|\hat{P}|\phi_0\rangle\right) \nonumber\\
    &= n_pn_q\dfrac{C[\N]\langle\N|\hat{P}|\phi_0\rangle}{J_{p,q}},
\end{align}
where $n_p$ and $n_q$ are occupation numbers and we have suppressed the spin indices for clarity. Thus, we get
\begin{equation}
  	\dfrac{\langle\N|\psi_{J_{p,q}}(\P)\rangle}{\langle\N|\psi(\P)\rangle} =  \dfrac{n_pn_q}{J_{p,q}}.
\end{equation}
Since the parameters in the reference ($\phi_i$) are complex, we need to consider derivatives with respect to their real and imaginary parts. For the derivative with respect to the real part, we get
\begin{align}
   \partial_{\text{Re}[\phi_i]}\langle\N|\psi(\P)\rangle =& \partial_{\text{Re}[\phi_i]}\left(C[\N]\langle\N|\hat{P}|\phi_0\rangle\right)\nonumber\\
   =&C[\N]\text{Re}\left[\partial_{\text{Re}[\phi_i]}\langle\N|\phi_0\rangle\right].
\end{align}
Similarly for the derivative with respect to the imaginary part
\begin{equation}
   \partial_{\text{Im}[\phi_i]}\langle\N|\psi(\P)\rangle  =C[\N]\text{Re}\left[\partial_{\text{Im}[\phi_i]}\langle\N|\phi_0\rangle\right].
\end{equation}
For Slater determinants
\begin{equation}
\begin{split}
   \text{Re}\left[\partial_{\text{Re}[\theta_{pi}]}\text{det}(\theta[\N])\right] &= n_p \text{Re}\left[\text{det}(\theta[\N])\theta[\N]^{-1}_{ip}\right],\\
   \text{Re}\left[\partial_{\text{Im}[\theta_{pi}]}\text{det}(\theta[\N])\right] &= -n_p \text{Im}\left[\text{det}(\theta[\N])\theta[\N]^{-1}_{ip}\right].\\
\end{split}
\end{equation}
For Pfaffians
\begin{equation}
\begin{split}
   \text{Re}\left[\partial_{\text{Re}[F_{pq}]}\text{pf}(F[\N])\right] &= \dfrac{n_pn_q}{2} \text{Re}\left[\text{pf}(F[\N])F[\N]^{-1}_{ip}\right],\\
   \text{Re}\left[\partial_{\text{Im}[F_{pq}]}\text{pf}(F[\N])\right] &= -\dfrac{n_pn_q}{2} \text{Im}\left[\text{pf}(F[\N])F[\N]^{-1}_{ip}\right].\\
\end{split}
\end{equation}

\section{Results}\label{sec:res}
Before discussing results we make a few remarks about notation for symmetry restored wavefunctions. All the reported VMC energies refer to wavefunctions that are $N$, $S_z$ and $K$ eigenfunctions. We use the prefix $S_z$ and K if these symmetries are broken and restored in the reference state. For example, K$S_z$GHF denotes a complex conjugation and $S_z$ projected GHF wavefunction, while $S_z$GHF denotes an $S_z$ projected GHF wavefunction which does not break the complex conjugation symmetry. We will denote number symmetry restored GBCS and RBCS wavefunctions by their conventional names Pfaffian and AGP, respectively. Jastrow factors are denoted by adding the prefix J to the wavefunction name.

The initial guess for Slater determinant calculations was computed using the Hartree-Fock modules in PySCF.\cite{sun2018pyscf} For pairing wavefunctions, we used the converged result of the corresponding Slater determinant calculations as the initial guess. We used our selected CI program Dice\cite{ShaHolUmr-JCTC-17,Holmes2016,smith2017cheap} to obtain full configuration interaction (FCI) energies and MOLPRO\cite{werner2012molpro} to get complete active space perturbation theory\cite{werner1996third, celani2000multireference} (CASPT2) energies for \ce{N2}. For several systems, we have also performed the fixed node Green's function Monte Carlo (GFMC) calculations.\cite{van1994fixed, Haaf1995} The GFMC calculations use the VMC wavefunctions as the trial state and deliver variational energies that are strictly between the VMC results and the true ground state energy. (%Any residual error in the GFMC calculations is solely due to the fixed-node approximation.%)
The details of the GFMC algorithm will be reported in a forthcoming publication.

\subsection{\ce{Hydrogen chain}}
\begin{figure}
\centering
\includegraphics[width=0.48\textwidth]{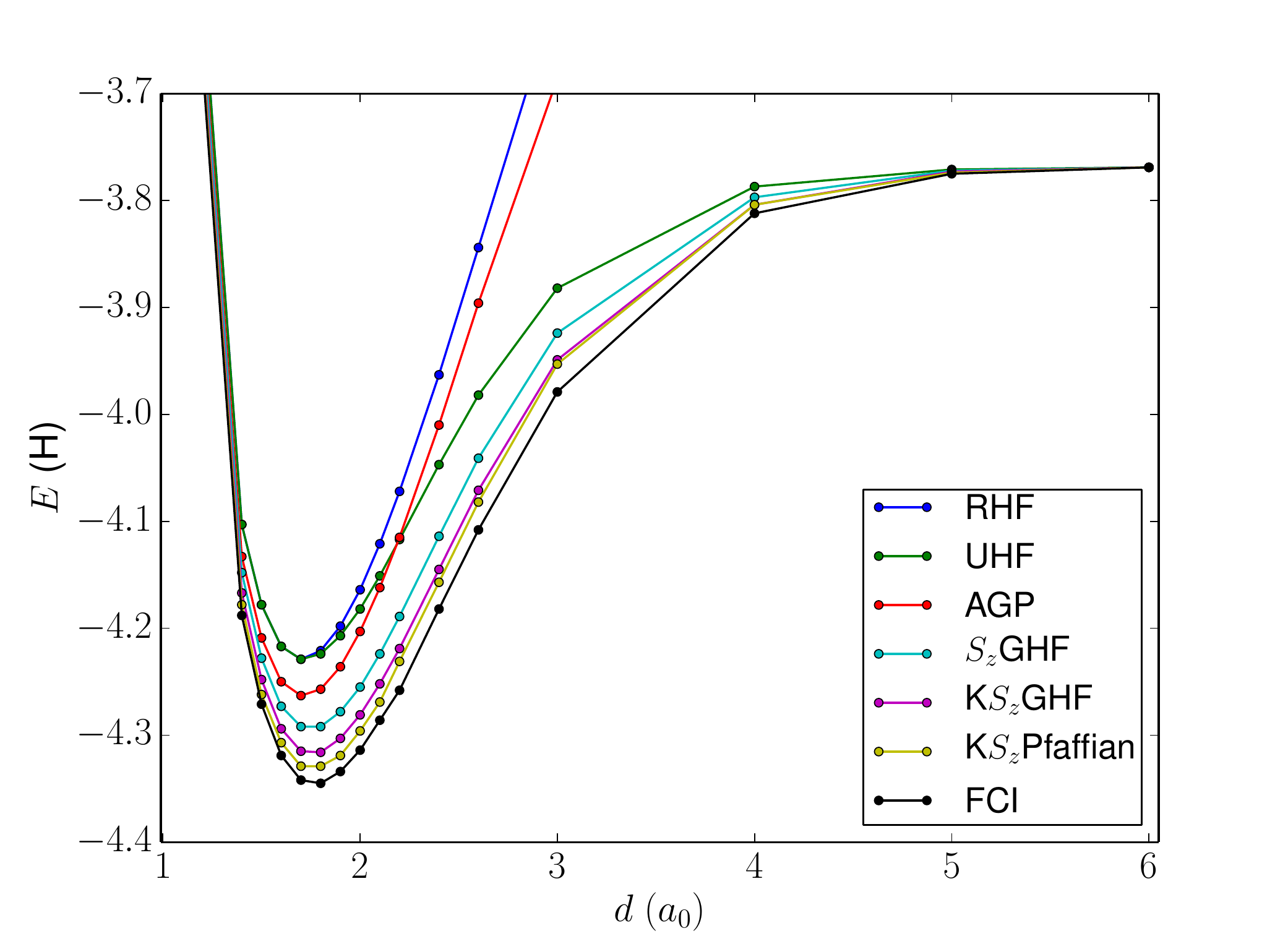}
\caption{\ce{H8} linear chain potential energy curves in the minimal sto-6g basis set. GHF energies are not shown because they are identical to UHF for this system. J-K$S_z$Pfaffian energies are not shown because they coincide with FCI energies on the scale of this plot.}\label{fig:h8}
\end{figure}
Hydrogen chains are important systems in their own right and embody some of the more interesting physics of real systems. They include long range Coulomb interactions and by changing the interatomic distance the strength of the correlation can be modulated. Exact results for large chains at all bond lengths can be obtained using the Density Matrix Renormalization Group (DMRG) algorithm. They are a challenging benchmark system for our method, particularly so, because our wavefunctions do not make use of the fact that the underlying system is one dimensional.

We first present results on the small open \ce{H8} chain using the minimal sto-6g basis to illustrate the quality of different wavefunctions discussed above. This system is small enough (4900 determinants in the $S_z=0$ subspace) to allow deterministic sampling (every determinant is visited once) of the wavefunction, so there are no stochastic errors in the results. Fig \ref{fig:h8} shows the ground state potential energy curves obtained for symmetric dissociation of the chain, and Table \ref{tab:h8} shows the errors in the energies relative to the FCI energy for three different geometries. The RHF and AGP wavefunctions are not size-consistent and do not approach the correct dissociation limit. All the other wavefunctions shown here converge to the exact FCI limit. For this system UHG and GHF energies are identical. Restoring the broken $S_z$ symmetry of the GHF wavefunction recovers slightly more than half of the missing correlation energy, while additionally breaking and restoring the complex conjugation symmetry cuts the remaining error by another 50\%. Breaking and restoring the number symmetry of the K$S_z$GHF wavefunction leading to K$S_z$Pfaffian results in further significant improvement in accuracy. The J-K$S_z$Pfaffian wavefunction (not shown in the plot) has energy errors less than 0.2 mE\textsubscript{h} for all the geometries considered. We have not restored the $S^2$ symmetry in any of these calculations, and work is currently underway to utilize this and other symmetries, e.g. point group symmetry, within our VMC implementation.
%It is often difficult to predict \emph{a priori}, breaking and restoring which symmetries will lead to large energy gains.

\begin{table}[htp]
\caption{Errors in the ground state energy (E\textsubscript{h}) for the \ce{H8} linear chain calculated using various wavefunctions.}\label{tab:h8}
\begin{tabular}{lllllll}
\hline
\hline
Wavefunction &~~~& $d=1.4$ &~~& $d=1.8$ &~~& $d=2.4$\\
\hline
RHF && 0.085 && 0.124 && 0.219 \\
UHF && 0.085 && 0.121 && 0.135 \\
AGP && 0.055 && 0.088 && 0.172 \\
$S_z$GHF && 0.040 && 0.053 && 0.068 \\
K$S_z$GHF && 0.021 && 0.029 && 0.037 \\
K$S_z$Pfaffian && 0.010 && 0.016 && 0.025 \\
J-K$S_z$Pfaffian && 0.0001 && 0.0002 && 0.0002 \\
\hline
\end{tabular}
\end{table}

\begin{table*}[htp]
\centering

\caption{Errors in energy (mE\textsubscript{h}) per electron for the ground state of the \ce{H50} chain with open boundary conditions as a function of interatomic distance (Bohr). Exact energies were obtained using DMRG. MC statistical errors are less than 0.02 mE\textsubscript{h}/electron.}\label{tab:h50}
\begin{tabular}{*{16}c}
\hline
\hline
$d $ &~~& \multicolumn{2}{c}{J-KRHF} &~~& \multicolumn{2}{c}{J-KAGP} &~~& \multicolumn{2}{c}{J-KUHF} &~~& \multicolumn{2}{c}{J-K$S_z$GHF} &~~& \multicolumn{2}{c}{J-K$S_z$Pfaffian}\\
\cline{3-4} \cline{6-7} \cline{9-10} \cline{12-13} \cline{15-16}
 & & VMC & GFMC && VMC & GFMC && VMC & GFMC && VMC & GFMC && VMC & GFMC\\
\hline
1.6 && 1.93 & 0.84 && 1.61 & 0.68 && 1.66 &	0.70 &&	0.92 & 0.30 &&	0.68 & 0.22\\
1.8 && 2.64 & 1.14 && 2.02 & 0.91 && 2.17 &	0.98 &&	0.94 & 0.26 &&	0.79 & 0.23\\
2.5 && 3.59 & 1.60 && 2.96 & 1.26 && 3.43 &	1.47 &&	0.76 & 0.18 &&	0.62 & 0.12\\
\hline
\end{tabular}
\end{table*}

Table \ref{tab:h50} shows the errors in ground state energies of an open \ce{H50} chain at different interatomic distances. We used the sto-6g minimal basis in this calculation. This much larger system has $\sim 10^{28}$ determinants in the $S_z=0$ subspace. For all geometries, we used the screening parameter value of $\epsilon = 10^{-6}$, which was sufficiently small to obtain results that are converged to all shown decimal places.\cite{Sabzevari18} The energy per electron obtained from the J-K$S_z$Pfaffian wavefunction is within 1 mE\textsubscript{h} of the exact DMRG results. The fixed node GFMC calculations performed using the converged VMC wavefunction as the trial wavefunction recover a significant amount of correlation while also giving a strictly variational result. The GFMC energies per electron obtained using the J-K$S_z$Pfaffian trial wavefunction are within 0.2 mE\textsubscript{h} of the DMRG energies. From these results, it is clear that Pfaffian wavefunctions seem to offer only a marginal improvement over GHF states in this case. This is in contrast to the results of the non-Jastrow calculations on the \ce{H8} chain where Pfaffian results were significantly superior to the GHF results, which indicates that the Jastrow factors are able to make up for the missing correlation between the Pfaffians and GHF in this system. Another important observation about the calculations is that the Pfaffian wavefunctions are significantly more difficult to optimize and many more iterations are needed to obtain (apparent) convergence. It is possible that more sophisticated, albeit expensive, optimization schemes such as %stochastic reconfiguration\cite{PhysRevB.64.024512} or
the linear method\cite{UmrTouFilSorHen-PRL-07,Toulouse2007,TouUmr-JCP-08} may converge to lower energies for these wavefunctions.

\subsection{\ce{N2}}
Correctly dissociating the N$_2$ molecule is a significant challenge for several electronic structure methods. Here we perform several SJMF calculations with various broken symmetries. The Jastrow factors are defined over orthogonal local orbitals and to obtain these orbitals we first symmetrically orthogonalized the atomic orbitals using Lowdin's ($\mathbf{S}^{-1/2}$) procedure. We performed an additional unitary transformation that performs rotations among orbitals on the same nitrogen atom to obtain $sp$ hybrid orbitals. In our testing, these hybrid orbitals were found to give lower energies and faster convergence compared to bare Lowdin orbitals.

\begin{figure}
\centering
\includegraphics[width=0.48\textwidth]{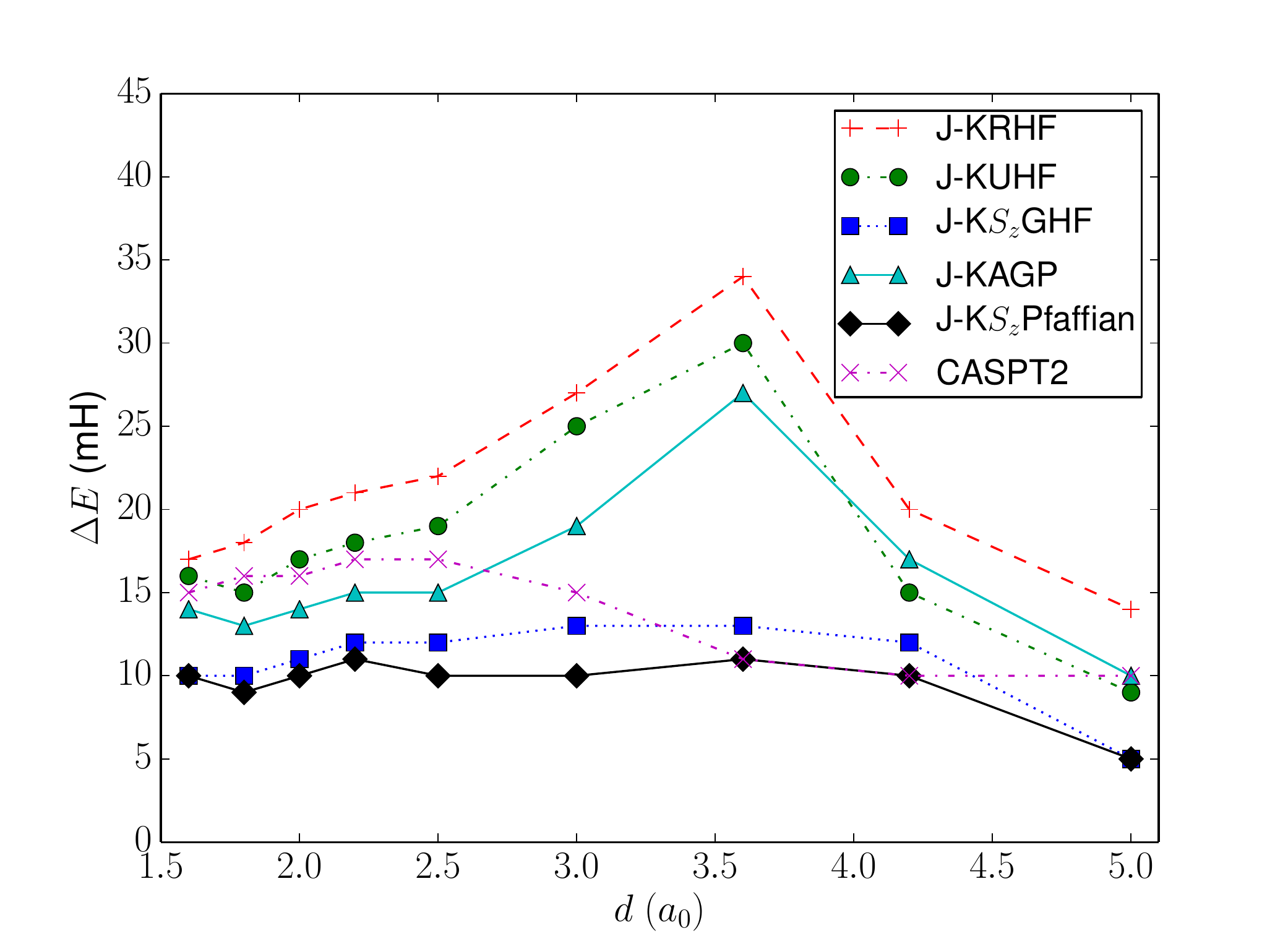}
\caption{\ce{N2} potential energy curve errors in the 6-31g basis set}\label{fig:n2}
\end{figure}

Fig. \ref{fig:n2} shows the errors in ground state energy in the 6-31g basis. The Jastrow-K$S_z$Pfaffian wavefunction gives better absolute energies than CASPT2, with $p$ valence active space (minimal active space required for qualitatively correct energies), for all geometries considered. It has a lower non-parallelity error as well. Although the Jastrow helps capture a significant amount of correlation in this basis set, our calculations with larger DZ/TZ basis sets have shown it to not be an efficient way of obtaining dynamical correlation. This suggests that a perturbation theory or CI expansions on top of these wavefunctions may be a better way to add dynamical correlation.

\subsection{Two dimensional tilted Hubbard model}
In this section, we consider the Hubbard model on a two dimensional tilted square lattice, with the Hamiltonian
\begin{equation}
    H = -t\sum_{\langle i, j \rangle, \sigma}\left(c^{\dagger}_{i\sigma}c^{\dagger}_{j\sigma}+c^{\dagger}_{j\sigma}c^{\dagger}_{i\sigma}\right) + U\sum_{i}n_{i\uparrow}n_{i\downarrow},
\end{equation}
where $\langle\ \rangle$ denotes nearest neighbors. %This model displays interesting behavior with strong correlations near half-filling and in the intermediate to large coupling range.
We report calculations on the $\ang{45}$ tilted $3\sqrt{2}\times3\sqrt{2}$ square lattice with periodic boundary conditions for which exact Lanczos diagonalization results are available.\cite{becca2000ground} Table \ref{tab:hubbard18} shows errors in the ground state energy at half-filling. It is again evident that breaking and restoration of complex conjugation symmetry lower the energy significantly.

\begin{table*}[htp]
\centering

\caption{Errors in ground state energy per electron (units of $10^{-3}t$) for the $3\sqrt{2}\times3\sqrt{2}$ tilted Hubbard model with periodic boundary conditions for different $U/t$ values at half-filling. MC statistical errors are less than $0.1\times 10^{-3}\ t/$electron.}\label{tab:hubbard18}
\begin{tabular}{*{15}c}
\hline
\hline
$U/t$ &~~& \multicolumn{2}{c}{J-$S_z$GHF} &~~& \multicolumn{2}{c}{J-K$S_z$GHF} &~~& \multicolumn{2}{c}{J-$S_z$Pfaffian} &~~& \multicolumn{2}{c}{J-K$S_z$Pfaffian}\\
\cline{3-4} \cline{6-7} \cline{9-10} \cline{12-13}
 & & VMC & GFMC && VMC & GFMC && VMC & GFMC && VMC & GFMC \\
\hline
4 && 7.3 & 2.6 && 5.1 &	1.9 && 6.8 & 2.3 &&	4.9 & 1.9 \\
8 && 10.2 & 4.1 && 7.0 & 2.0 &&	10.1 & 3.9 && 6.8 &	2.0\\
10 && 8.9 & 2.5 && 5.7 & 1.4 &&	7.6 & 2.3 && 5.4 & 1.2\\
20 && 3.9 & 0.7 && 2.9 & 0.2 &&	3.9 & 0.6 && 2.9 & 0.2\\
\hline
\end{tabular}
\end{table*}

In table \ref{tab:hubbard98}, we show the results for the much bigger half-filled $7\sqrt{2}\times7\sqrt{2}$ lattice containing 98 sites. Since exact energies for this lattice are not available, we compare our energies to GFMC energies reported in reference \citenum{LeBlanc2015}. These were obtained using a Jastrow-Slater trial wavefunction with backflow correlation included and were shown to converge to a thermodynamic limit with an error of $0.0015t$ relative to the exact AFQMC limit.

\begin{table}[htp]
\caption{Ground state energy per electron (units of $t$) for the $7\sqrt{2}\times7\sqrt{2}$ tilted Hubbard model with periodic boundary conditions for different $U$ values. The energies in the first two columns use the J-K$S_z$GHF wavefunction. The E$_{\text{ref}}$ results are taken from reference \citenum{LeBlanc2015}. MC statistical errors are less than $0.1\times 10^{-3}\ t/$electron.}\label{tab:hubbard98}
\begin{tabular}{ccccccc}
\hline
\hline
$U/t$&~~~~&VMC&~~~&GFMC&~~~&E$_{\text{ref}}$\\
\hline
2&& -1.1920 && -1.1939 && -1.1962\\
4&& -0.8566 && -0.8598 && -0.8620\\
8&& -0.5183 && -0.5221 && -0.5237\\
\hline
\end{tabular}
\end{table}

Correlation functions can be used to extract useful physical information from a wavefunction. Their accuracy reflects the quality of the wavefunction. Here we calculate the density-density correlation functions given by
\begin{equation}
    N(i,j) = \dfrac{\langle\psi|n_in_j|\psi\rangle}{\langle\psi|\psi\rangle}.
\end{equation}
This function can be calculated using Monte Carlo sampling in a manner analogous to energy and gradient calculations:
\begin{align}
N(i,j) &= \frac{\sum_{\N} |\langle \Psi(\P) |\N \rangle|^2 \frac{\langle \N|n_in_j|\Psi(\P)\rangle}{\langle \N|\Psi(\P)\rangle}}{\sum_{\N} |\langle \Psi(\P)|\N\rangle|^2}\nonumber\\
&= \sum_{\N} \rho_{\N} N^{ij}_L[\N] = \langle N^{ij}_L[\N]\rangle_{\rho_{\N}},
\end{align}
where, $N^{ij}_L[\N] = \frac{\langle \N|n_in_j|\Psi(\P)\rangle}{\langle \N|\Psi(\P)\rangle}$ is the local correlation function that is averaged during a Monte Carlo run, and $i$ and $j$ are spatial orbital indices. Note that, unlike the energy, local correlation function does not satisfy the zero variance principle and we expect the results to be noisier than the energies. In table \ref{tab:corr}, the values of the correlation function are shown for the $3\sqrt{2}\times3\sqrt{2}$ lattice with $U/t=4$. In this lattice, only five unique $\langle n_i n_j\rangle$ values exist due to symmetry. For reference we use the values obtained using DMRG, which for this small two-dimensional system, gives correlation function values very close to that of the exact wavefunction. The agreement between the VMC and DMRG wavefunctions is good and is not worse than the error in the total energies. %At this intermediate value of $U$, spin contamination in the GHF and Pfaffian wavefunctions is not very severe, which can be confirmed from the homogeneity of spin resolved charge densities. Despite this, the small differences in the correlation function will likely vanish if the $S^2$ symmetry is restored.

\begin{table}[htp]
\caption{Density correlation function $\langle n_i n_j\rangle$ for the $3\sqrt{2}\times3\sqrt{2}$ tilted Hubbard model with periodic boundary conditions for $U/t=4$ at half-filling for different values of distance between sites $i$ and $j$. MC statistical errors are less than $0.001$.}\label{tab:corr}
\begin{tabular}{ccccccccccc}
\hline
\hline
Wavefunction &~~~& $d=1$ &~~& $d=\sqrt{2}$ &~~& $d=2$ &~~& $d=\sqrt{5}$ &~~& $d=3$\\
\hline
DMRG && 0.944 && 0.993 && 0.992 && 0.991 && 0.998\\
J-K$S_z$GHF && 0.942 && 0.992 && 0.993 && 0.992 && 0.997\\
J-K$S_z$Pfaffian && 0.941 && 0.993 && 0.992 && 0.990 && 0.997\\
\hline
\end{tabular}
\end{table}

\section{Conclusions}\label{sec:conc}
In this paper, we have presented a VMC algorithm for optimizing SJMF wavefunctions. We described a unified hierarchy of wavefunctions that have appeared in different contexts. VMC provides an efficient route to optimizing these wavefunctions and the symmetry projectors used here can be applied in a natural and efficient manner. Our benchmark calculations demonstrate that these wavefunctions are capable of accurately describing strong correlations. In particular, the restoration of complex conjugation symmetry, which has not been used in VMC before to the best of our knowledge, yields significant relaxation of energies and can be potentially used with many other wavefunctions. Because Jastrows are capable of local number and $S_z$ projections, the J-$S_z$Pfaffian and J-$S_z$GHF wavefunctions are exactly size-consistent.

Other symmetries, including $S^2$, point group, time reversal, and translational symmetry can also be restored in a VMC approach and work is underway in this direction. This will allow us to obtain more accurate correlation functions. It will be a challenge to make $S^2$ projected wavefunctions size-consistent, while also keeping the calculation cost down. Another possible improvement is using more sophisticated optimization methods such as the linear method to avoid the large number of iterations needed to optimize the wavefunctions containing Pfaffians. We are also working towards implementing ways to add dynamical correlation beyond these wavefunctions using the configuration interaction approach and perturbation theory.\cite{ci-mps,guillaume17}

Since the accuracy and performance of projection Monte Carlo techniques depends critically on the quality of the trial wavefunction used, our VMC wavefunctions will be useful in such approaches. It will be interesting to analyze how symmetry breaking and projection affect the nodal structure of the wavefunction. Using the approach developed in Ref. \citenum{chang2016auxiliary}, one can use Jastrow correlated wavefunctions in AFQMC as well.

\section{Acknowledgements}
We thank Cyrus Umrigar for several helpful discussions. The funding for this project was provided by the national science foundation through the grant CHE-1800584.
\providecommand{\latin}[1]{#1}
\makeatletter
\providecommand{\doi}
  {\begingroup\let\do\@makeother\dospecials
  \catcode`\{=1 \catcode`\}=2 \doi@aux}
\providecommand{\doi@aux}[1]{\endgroup\texttt{#1}}
\makeatother
\providecommand*\mcitethebibliography{\thebibliography}
\csname @ifundefined\endcsname{endmcitethebibliography}
  {\let\endmcitethebibliography\endthebibliography}{}

\end{document}